\documentclass[journal]{IEEEtran}

\usepackage[ruled]{algorithm2e}
\usepackage{eso-pic}
\usepackage{graphicx}
\usepackage{tikz}
\usepackage{pgf}
\usepackage{lscape}
\usepackage{amssymb}
\usepackage{fancyvrb}
\usepackage{amsfonts}
\usepackage{mathtools}
\usepackage{verbatim}
\usepackage{algorithm2e}
\usepackage{multirow}
\usepackage{proof}
\usepackage{wrapfig}
\usetikzlibrary{arrows,automata}
\usetikzlibrary{shapes}
\usepackage{epstopdf}
\usepackage{xfrac}
\usepackage{url}
\usepackage[american]{babel}
\usepackage{booktabs}
\usepackage{makecell}

\usepackage[square,sort,comma,numbers]{natbib}

\begin{document}

\title{Blockchain for the Internet of Things: \\Present and Future}

 \author{Francesco~Restuccia,~\IEEEmembership{Member,~IEEE}, Salvatore~D'Oro,~\IEEEmembership{Member,~IEEE}, Salil S. Kanhere,~\IEEEmembership{Senior Member,~IEEE,}~Tommaso~Melodia,~\IEEEmembership{Fellow,~IEEE}, and Sajal K. Das,~\IEEEmembership{Fellow,~IEEE} 
\thanks{F. Restuccia, S. D'Oro and T. Melodia are with the Institute for the Wireless Internet of Things, Department of Electrical and Computer Engineering, Northeastern University, Boston,
MA, 02115 USA e-mail: \{frestuc, salvatoredoro, melodia\}@northeastern.edu. \newline S.S. Kanhere is with the School of Computer Science and Engineering, University of New South Wales, Sydney, NSW 2052, Australia. E-mail: salil.kanhere@unsw.edu.au. \newline S. K. Das is with the Department of Computer Science, Missouri University of Science and Technology, Rolla, MO 65401 USA. Email: sdas@mst.edu.
}
\thanks{Manuscript received December 15, 2017; revised January 1, 2018.}
}

\markboth{IEEE Internet of Things Journal,~Vol.~1, No.~1, January~2018}%
{Shell \MakeLowercase{\textit{et al.}}: Bare Demo of IEEEtran.cls for IEEE Journals}

\maketitle

\begin{abstract}
 One of the key challenges to the IoT's success is how to secure and anonymize billions of IoT transactions and devices per day, an issue that still lingers despite significant research efforts over the last few years. On the other hand, technologies based on blockchain algorithms are disrupting today's cryptocurrency markets and showing tremendous potential, since they provide a distributed transaction ledger that cannot be tampered with or controlled by a single entity. Although the blockchain may present itself as a cure-all for the IoT's security and privacy challenges, significant research efforts still need to be put forth to adapt the computation-intensive blockchain algorithms to the stringent energy and processing constraints of today's IoT devices. In this paper, we provide an overview of existing literature on the topic of blockchain for IoT, and present a roadmap of research challenges that will need to be addressed to enable the usage of blockchain technologies in the IoT.  
\end{abstract}


\IEEEpeerreviewmaketitle

\section{Introduction}

It is hard to mention a technology that will impact and benefit our lives more than the Internet of Things (IoT). In a few years, cars, kitchen appliances, televisions, smartphones, utility meters, intra-body sensors, thermostats, and almost anything we can imagine will be  absorbed into the Internet and accessible from anywhere on the planet \cite{whitmore2015internet}. The revolution brought by the IoT will be unmatched -- some say it will be similar to the building of roads and railroads which powered the Industrial Revolution of the 18th to 19th centuries \cite{ForbesIoT} -- and will take by storm every human sector and industry, ranging from education, health-care, smart home and smart city, to manufacturing, mining, commerce, transportation, and surveillance, just to mention a few \cite{da2014internet}.

Over the last few years, researchers have mainly focused their attention on addressing IoT's computation and communication scalability issues \cite{bello2013communication,tayeb2017survey,gubbi2013internet,restuccia2012performance,restuccia2016optimizing,jagannath2019machine}. While these topics are certainly paramount to IoT's success and need to be thoroughly investigated, the community has now widely acknowledged that they have to be considered ``low-hanging fruits'' with respect to the towering issues of IoT security and privacy, which are unprecedented in scope and magnitude \cite{restuccia2018securing,restuccia2017quality,bekara2014security,frustaci2017evaluating,alrawais2017fog,stojmenovic2014fog,Zhang-cns2017,Zhang-tmc2018} and will require considerable research effort to be overcome. It is easy to imagine, indeed, that once humans, sensors, cars, robots, and drones are able to seamlessly interact with each other from any side of the globe, a number of threats that we cannot even imagine today will be unveiled.  

As currently envisioned, the IoT will implement a centralized, client-server based access model in which IoT transactions (\textit{i.e.}, data, money, or any other object of value) between IoT entities (\textit{i.e.}, any computing device or stakeholder connected to the IoT) is entrusted to monolithic, centralized service providers \cite{ali2017iot}. This model clearly simplifies the interactions between IoT entities and facilitates the data collection process. However, it ultimately makes the IoT vulnerable to a number of spinous security and privacy issues. Specifically, centralized service providers can make illegitimate use of IoT data, for example, mass-surveillance programs \cite{GuardianSpy}. Even more importantly, centralized data collection models can expose the system to hacking by malicious activities, with nefarious consequences for citizens, as unveiled in \cite{BabyIoTSecurity,UnlockDoorsIoT,PopScience,PaceMakerDeath}. Another major challenge is the authentication of IoT entities that will be mostly deployed \emph{in the wild} with little supervision \cite{kalra2015secure,amin2018light}. If not addressed, IoT authentication issues can generate botnets (\textit{e.g.}, Mirai \cite{kolias2017ddos}) and hard-to-tackle sybil attacks \cite{zhang2014sybil}. 
The key intuition to address the challenges above is to \textit{orchestrate IoT transactions in a decentralized fashion}, so that no single entity has control over them. Not only will decentralization provide security and privacy by design, but also empower users with the choice of sharing or selling their sensor data with third party entities without intermediaries. Decentralized control also implies scalability -- which has plagued the IoT from its very inception \cite{gharbieh2017spatiotemporal,miorandi2012internet}. The end goal, therefore, is to investigate decentralized data access models for the IoT, which will ensure that user-data is not entrusted to centralized entities or companies, but instead is made the property of the users themselves. To this end, technologies and systems based on the concept of \textit{blockchain} have enabled the cryptocurrency market, and may prove crucial to achieve the stringent security and privacy goals of the IoT \cite{conoscenti2016blockchain}. Although the key algorithms and principles behind the blockchain have been known since the 70's (\textit{i.e.}, Merkle trees \cite{merkle1980protocols}, consensus algorithms \cite{Zheng-ieeebigdata2017}), the first practical application of the blockchain was originally proposed in 2008 as part of the Bitcoin cryptocurrency \cite{nakamoto2008bitcoin}. Since then, it has been widely applied to a wide range of non-monetary applications, including transportation, energy management, smart cities, drones/robots, and manufacturing; we survey existing literature on the topic in Section \ref{sec:iot_block_app}. 


In a nutshell, a blockchain maintains a collection (or \emph{ledger}) of transactions in a decentralized fashion -- we describe in details what a blockchain is in Section \ref{sec:blockchain}. The ledger is \emph{immutable}, meaning that past transactions cannot be modified by any entity registering transactions in the blockchain\footnote{Many works refer to the blockchain as an immutable data structure, however it is technically imprecise to define it as immutable. In fact,
there are precedents where entries in the blockchain have been changed after attacks or misbehavior of the network \cite{werbach2017trust}. In this paper, the word immutable is intended to be used to represent the \textit{hard-to-change} structure of the blockchain \cite{walch2016path}.},  and is shared and synchronized across all participating nodes. This way, the blockchain guarantees that the ledger cannot be tampered with, and that all the data held by the blockchain is trustworthy. A \textit{consensus algorithm}, which involves solving a hard-to-solve (\textit{i.e.}, resource-demanding) yet easy-to-verify puzzle called proof-of-work (PoW), is used for appending (\textit{mining}) new blocks into the blockchain, and thus establish a {\em secure trusted network among untrusted entities}. For identification purposes, blockchain nodes may choose to employ changeable public keys to prevent tracking. Multiple transactions are merged together to form a block which is appended to the ledger by following the consensus algorithm. Each block includes the hash of the previous block in the ledger (hence the name \emph{blockchain}). Any modifications to a block (and thus transactions) can be readily detected as the hash maintained in the subsequent block will not match.

The combination of blockchain and IoT has disruptive potential. Indeed, the blockchain may help the IoT's expansion into our society by providing the following key advantages:
\begin{itemize}
\item \textit{Anonymity.} IoT entities can participate to the blockchain with a public/private key, which (if so desired) does not reveal in itself the real identity of the entity;
\item \textit{Decentralization.} Traditional centralized systems require each transaction be validated through a centralized authority (\textit{e.g.}, a central bank) -- which inevitably translates into a performance bottleneck. Conversely, third-party validation is no longer needed in the blockchain, since consensus algorithms maintain data consistency;
\item \textit{Non-repudiation.} The blockchain ensures that (i) transactions can be easily validated; and (ii) invalid transactions are not admitted -- it is nearly impossible to delete or roll back transactions once included in the blockchain.
\end{itemize}

Although the blockchain may look as a panacea to the IoT's security and privacy issues, there are still many research challenges that prevent its off-the-shelf application to most of today's IoT networks. Indeed, most of the algorithms used by today's blockchain-based systems were not designed to be run on devices with extremely stringent computation/energy/bandwidth constraints as in the IoT.  Several key challenges (discussed in detail in Section \ref{sec:res_challenges}) need to be addressed, including: (i) scalability issues that stem from the need to achieve consensus among potentially billions of miners; (ii) high computation demands due to the use of
proof-of-work (or similar) algorithms; and (iii) high delays due to anti-double spending mechanisms (issue which may not necessarily apply to the IoT).

The focus of this paper is to provide an overview of the state of the art  pertaining to the application of blockchain-based system to address IoT's security and privacy issues, and offer a roadmap of novel and exciting research challenges to the research community. We point out that an in-depth survey and comparison of existing blockchain-based IoT systems is not the ultimate objective of this paper. Instead, our main goal is to prime the readers and stimulate their research efforts toward the development of next-generation secure-by-design blockchain-based IoT systems.

\section{What is a Blockchain?}\label{sec:blockchain}

\begin{figure}[h!]
  \centering
    \includegraphics[width=\columnwidth]{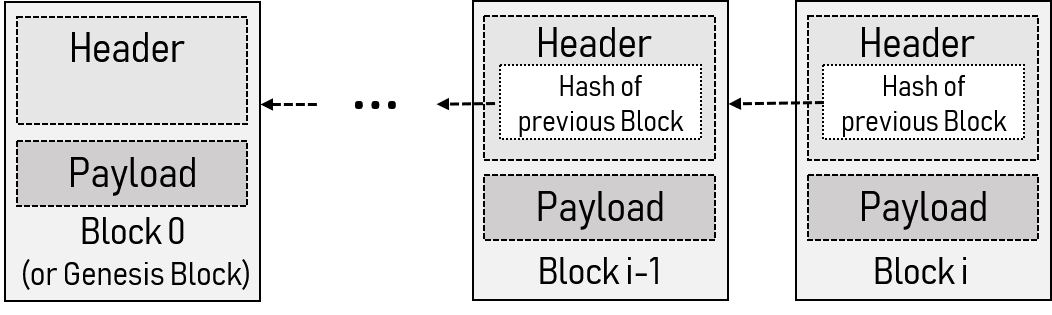} 
    \caption{The structure of a blockchain.}
    \label{fig:blockchain}
\end{figure}

From a computational viewpoint, a \textit{blockchain} is a data structure where entries (also called \textit{blocks}) are stored and linked to one another in sequential order. 
As shown in Fig. \ref{fig:blockchain}, the concept of blockchain is very similar to that of a linked list, where each entry is linked to the next one by means of a pointer. Although the two structures above are conceptually the same, their implementation differs in several major aspects.

Each block is composed by a header and a data payload. While the payload is generally used to store a list of transactions among users of the blockchain, the header is used to convey useful information with respect to the block, such as its length and content. Furthermore, the header stores the 32-bit SHA256 \cite{pub2012secure} hash value of the previous block. The importance of such a field is twofold: this way, (i) each block is immutably linked to the previous one; and (ii) the hash value of the $i$-th block will depend on the hash value of block $i-1$. The first feature provides a very efficient mechanism to interconnect all blocks of the blockchain, while the second one,  as discussed later, is used to prevent malicious attacks. 

To understand this latter statement, it is important to first understand how the blockchain is generated and maintained over time.

\subsection{Consensus Mechanisms} The purpose of the blockchain \cite{nakamoto2008bitcoin} is to enable peer-to-peer transactions that are validated, organized into blocks, and then stored inside a distributed ledger. To achieve this objective, the blockchain is regulated by decentralized consensus algorithms that determine how and when a group of transactions can be included in the ledger. Specifically, each new block can be appended to the blockchain only if the majority of nodes in the network agrees upon its inclusions, that is, only if consensus among users of the blockchain is reached. 
Each node in the network keeps a local copy of the blockchain. When a new block achieves a consensus, it is broadcast over the network. Thus, each node appends the new block to its local copy of the blockchain. These mechanisms make it possible to create multiple consistent copies of the blockchain, such that as soon as the majority of nodes possesses the same copy of the blockchain, the network can be considered as reliable and trusted.

Consensus is an extremely important concept for the blockchain. The first implementations of the blockchain adopted the proof-of-work (PoW) consensus mechanism \cite{dwork1992pricing}, which provides a distributed mechanism to maintain and validate the blockchain. The idea behind PoW is to achieve consensus among nodes of the network through hard-to-compute, but easy-to-verify, computational puzzles. For example, the Bitcoin blockchain asks its users (also called \textit{miners}) to find a $4$-byte random number, \textit{i.e.}, the \textit{nonce}, such that the SHA256 hash value of a new block is equal to or less than a given threshold. While the computation of the nonce is hard and computational-hungry, verifying that a nonce satisfies the threshold requirement is computationally very inexpensive. Accordingly, the first node that finds a candidate nonce notifies the blockchain network and broadcasts the new block. The obtained nonce, which represents the PoW of the miner, is tested by other nodes that determine whether or not the nonce is an actual solution of the hashing puzzle. When the validation of the nonce is successful, nodes add the new block to their locally stored blockchain and start working on a new block. 

\subsection{Computational Aspects} 

Although PoW is a very effective mechanism to achieve consensus, it requires overwhelming amounts of energy and computational power, which increase every year as more and more miners and transactions add up to the blockchain \cite{o2014bitcoin}. For this reason, other consensus mechanisms \cite{bach2018comparative} have been considered in many blockchain architectures. As an example, Proof-of-Stake (PoS) mechanisms \cite{kiayias2016provably,zheng2017overview} use deterministic rules based on the amount of coins, \textit{i.e.}, the \textit{stake}, to select which node in the network will append the next block to the blockchain. Similarly, the proof-of-importance (PoI) considers the stake as an important metric as well -- however, it  accounts also for metrics that measure the miner's involvement in the network, such as number and volume of transactions.

As shown in Fig. \ref{fig:blockchain}, the hash value of each block depends on the hash value of the previous blocks. Thus, a change in any of the already existing blocks of the blockchain would produce a different hash value for that block, that will then generate a cascading effect on all subsequent blocks and their hash values. The newly generated hash values will be different from those already stored by all other nodes in the network, and thanks to the consensus algorithm, the corrupted blocks will be rejected from the blockchain. 

\subsection{Security Aspects} In general, consensus algorithms guarantee the trustworthiness of a blockchain. However, there are cases where malicious users can leverage the blockchain structure to change, duplicate or delete blocks \cite{bradbury2013problem}. Specifically, it is sufficient for an attacker to possess more than the $50\%$ of the nodes in the network to take control of the whole blockchain. This attack, also referred to as the \textit{$51\%$ attack}, aims at governing the consensus mechanism to manipulate a blockchain. These attacks have been shown to be effective against many minor crypto-currencies such as Verge, Bitcoin Gold and Zencash \cite{51attack1} -- however, they have as well threatened even widespread crypto-currencies such as Bitcoin \cite{51attack2,51attack3}. 
Double-spending \cite{karame2012double,lin2017survey}, which consists in the replication of one or more transactions, is the main purpose of $51\%$ attacks. However, it has been shown that double-spending can be achieved even without approaching the $50\%$ threshold \cite{karame2015misbehavior}. 

To mitigate $51\%$ attacks, ever more blockchain-based systems are adopting better security strategies. For example, real-time validators can be used to increase the attack threshold to $99\%$ \cite{99percent}, which means that an attacker can take control over the blockchain network only if it has access to almost all nodes in the network. Another approach is  to use PoS consensus mechanisms where the importance given to the possession of coins (rather than of computational power) makes the $51\%$ attack unprofitable for the attacker and less likely to happen. 

\section{Overview of Blockchain-based IoT Systems}\label{sec:iot_block_app}

In this section, we provide a survey of the most relevant blockchain-based IoT systems investigated so far in literature. We divide the papers by categories, each named after the most common IoT applications nowadays available, \textit{i.e.}, smart energy, smart environments, robotics, transportation, and supply chain.


\begin{itemize}
\item \emph{Smart Energy.} This field has attracted  significant attention from the IoT community over the last years \cite{dalipi2016security}. The majority of the proposed IoT systems leverage the blockchain to (i) preserve the privacy of the users along with their personal information; and (ii) protect the system from malicious transaction such as users attempting to sell or buy unreasonable amount of energy \cite{laszka2017providing,lombardi2018blockchain,mylrea2017blockchain}. The authors in \cite{hahn2017smart,mengelkamp2018blockchain} propose auction systems where
users can sell to the highest bidder their excessive energy based on an auction defined in a smart contract, hence eliminating the need for a third-party auctioneer. Moreover, Hahn \emph{et al.} \cite{hahn2017smart} implemented the auction on a campus-level energy grid. Yan \textit{et al.} \cite{yan2017distributed} explored the use of blockchain to reconstruct the current distributed energy transaction patterns to allow decentralized real-time transactions and intelligent energy trading contracts using an automatic trust mechanism.
\item \emph{Smart Environments.} Smart environments have been extensively in industrial settings \cite{schaffers2011smart}, for smart healthcare \cite{catarinucci2015iot}, smart cities \cite{zanella2014internet} and smart homes \cite{kelly2013towards,dorri2017lsb,palai2018empowering}. In this context, the blockchain is used to ensure the availability and unrepudiability of sensed data collected in the wild, \textit{e.g.}, a farm area \cite{ibba2017citysense,patil2017framework}.
\item \emph{Robotics.} Existing work in this area leverages the blockchain as a system to support secure and reliable unmanned air vehicles (UAV) communications. Indeed, UAVs need to reliably coordinate their actions, exchange data and collaboratively make decisions. Sharma \emph{et al.} \cite{sharma2017socializing} present a system where  drones are programmed to use the blockchain to securely relay information. Moreover, Ferrer \emph{et al.} \cite{ferrer2016blockchain} investigate the use of the blockchain to provide security, autonomy and collective decision-making in swarm robotic systems. The authors in \cite{liang2017towards} leverage a combination of blockchain and cloud storage to protect the integrity of drone-collected data.
\item \emph{Transportation.} Over the last years, many IoT concepts have been used to design next-generation transportation systems \cite{liu2012vehicular,he2014developing,gerla2014internet}. The most promising aspect is that smart vehicles will likely not be as computationally constrained as other IoT devices, such as sensor platforms. Therefore, the blockchain is a strong candidate to become a system for tamper-proof data exchange among smart vehicles, as proposed by Steger \emph{et al.} \cite{steger2018secure}. Similarly, Ceba \emph{et al.} \cite{cebe2018block4forensic} monitor vehicle-related data (\textit{e.g.}, maintenance information and vehicle diagnosis
reports) by using the blockchain.  Yuan \emph{et al.} \cite{yuan2016towards} use the blockchain to design a full-fledged intelligent transportation system architecture, which includes application, contract, incentive, consensus, data, physical, and network layers. The blockchain has also been leveraged to implement systems to handle the public keys of the vehicles \cite{rowan2017securing}, and in general share data without third-party centralized management \cite{leiding2016self,sharma2017block}.  Li \emph{et al.} \cite{li2018creditcoin} propose \emph{CreditCoin}, a privacy-preserving system to share relevant information (\textit{e.g.}, accident, traffic) between vehicles, where participants are rewarded through monetary tokens. Yang \textit{et al.} \cite{yang2017blockchain} proposed a blockchain-based reputation system that estimates the trustworthiness of received messages.
\item \emph{Supply Chain \& Others.} Some systems have been proposed to enhance the functionality of cloud-based and on-demand manufacturing \cite{bahga2016blockchain,rabah2017overview}. A blockchain-based distribution framework to share knowledge and services across enterprises is presented in \cite{li2018toward}. A set of papers \cite{boudguiga2017towards,lee2017blockchain,samaniego2016hosting,samaniego2017virtual,stanciu2017blockchain} address edge computing, virtualization of IoT resources, among others.
\end{itemize}

\section{Blockchain technologies for the IoT} \label{sec:techsforiot}

In this section, we discuss the most important blockchain technologies and features, and we discuss their application to the IoT.\vspace{-0.3cm}

\subsection{Smart Contracts}

One of the key challenges of the IoT is to enable and control autonomous and self-organized machine-to-machine (M2M) communications. 
In this specific context, it is of paramount importance to design management mechanisms such that (i) interactions are automatically initiated; and (ii) there is no need to individually control and verify the trustworthiness of each interaction/communication. 
The above problem is definitely not trivial, and its complexity is further exacerbated by the large number of connected devices and their heterogeneous design. 
It is worth mentioning that the above problem is not peculiar of the IoT only, but it also affects all of those network architectures and systems where the lack of centralized entities that perform centralized network control and management call for self-organized and automated protocols. 

The best example is the blockchain, a system where distributed entities are required to autonomously reach consensus by locally executing complex algorithms.
In this context, \textit{smart contracts} \cite{christidis2016blockchains} has been shown to be effective to solve the above challenges.

In a nutshell, smart contracts are software programs that specify and automatically enforce contracts among two or more parties. To understand how smart contracts work, we consider the case where Alice rents a house to Bob. Bob is required to send a monthly payment to Alice. 
In the context of the blockchain, the above transaction can be easily encoded into a smart contract. 
As an example, in Ethereum's blockchain each smart contract is represented by a series of computational operations that are expressed via a programming language that is specified by an Application Binary Interface (ABI).
Indeed, it is sufficient to write a few lines of code to generate and link the contract to Bob, such that the monthly payment can be automatically triggered by a software program when the monthly deadline is over.
Therefore, smart contracts implement effective mechanisms to send/receive payments (i.e., the rent) to/from other entities when one or more conditions (i.e., monthly deadline) are satisfied. 

Although the previous example is very simple, contracts can generally implement very complex operations and can also be linked one to another, thus generating a nested structure (\textit{e.g.}, sublease). The advantages of smart contracts are numerous, and their impact on IoT networks is considerable, as discussed in \cite{christidis2016blockchains}. First, since contracts are stored inside the blockchain, their content is trusted among parties as it cannot be modified or corrupted after its inclusion in the blockchain. Second, each contract is assigned an unequivocal address in the blockchain and can be directly accessed from the Internet, thus making smart contracts well-suited to be accessed by remotely connected IoT devices. Finally, contracts consist of few lines of code that devices can easily understand and execute.

Given the similarities between the IoT and the blockchain, and looking at the success and effectiveness of smart contracts in blockchain applications, it is reasonable to assume that smart contracts can find useful applications in the IoT to support autonomous and self-organized interactions.
Although the application of smart contracts to the IoT is still being investigated, preliminary results already show that several IoT applications would benefit from blockchain technologies such as smart contracts. For example, the application of smart contracts to devise access control mechanisms that regulate the access to the IoT network has been shown to be beneficial for the IoT \cite{lin2018secure,ouaddah2016fairaccess,8306880,xu2018blendcac,zhang2018smart}. These works leverage the immutability of the blockchain to generate real-time access control lists that also regulate and describe access policies to device resources.
Another example is the work in \cite{christidis2016blockchains}, where authors discuss the possibility to achieve smart supply chain monitoring by means of smart contracts. They show that, not only smart contracts can be used to regulate transactions and fees related to the production and shipment processes of goods, but they can also be used to keep track of their position.

\subsection{Software and Content Validation}

The IoT system is well-known to be a heterogeneous environment where substantially different devices (in terms of hardware and offered services) interact with both users and other devices. In this challenging scenario, it is crucial to guarantee that the software embedded in each device (\textit{e.g.,} firmware, scripts)  is always up-to-date and satisfies regulations and security requirements of the network. Although the sheer number of devices in the network complicates the design of mechanisms that meet the above requirements in large-scale heterogeneous networks, the blockchain already provides useful embedded features that completely, or partially, address the above issues. 

As shown in \cite{lee2017blockchain,kouzinopoulos2018using}, the distributed nature of the blockchain may be leveraged to store and disseminate secure and verified firmware updates over the network. Specifically, the blockchain can be used to (i) store either the firmware update itself, or the address of a safe and trusted location where the updated code can be downloaded and installed; and (ii) use PoW (or similar tools) to determine when a device possesses an updated and verified firmware, thus  deciding whether or not a device can be trusted. Since the blockchain is maintained through consensus mechanisms, it is possible to generate trusted blockchains that store all trusted and up-to-date firmware updates \cite{lee2017blockchain} that can be easily identified and downloaded by network nodes. 

Another interesting application of blockchain technologies to IoT systems is the possibility to provide reliable license validation tools to avoid piracy and preserve copyrights of software/hardware developers \cite{herbert2015novel} and content creators \cite{poet}. Indeed, IoT devices are nowadays capable of performing heterogeneous sensing and computational tasks, and can be reprogrammed by dynamically loading different software applications coming from different developers. Although open-source software is now widely used in many IoT environments, there are still several applications whose code can be purchased from the Internet through licensing. The purpose of \cite{herbert2015novel} is to leverage blockchain technologies to provide effective licensing validation tools for a software developer to enforce their copyright.

\section{The Road Ahead}\label{sec:res_challenges}

We now propose a roadmap of research challenges pertaining to the application of blockchain algorithms to the IoT.

\subsection{Addressing Blockchain Scalability Issues}

Applying blockchain technologies to the IoT implies that scalability issues must be addressed. Most importantly, existing blockchain tools require all nodes in the network to either approve each transaction/block, or store them locally. While these tasks are easy for personal computers or workstations, they may be prohibitive for small sensors with limited storage and computational resources. This issue is further exacerbated by the fact that the amount of data transmissions, and thus required transactions, to be stored in the blockchain is large and exponentially increasing over time \cite{7946872,conoscenti2016blockchain,worner2014your,barber2012bitter}. In other words, existing consensus algorithms that rely on PoW and PoS are not directly applicable  to address long-term, reliable and scalable solutions for blockchain-based IoT systems. 


The most widely used approach to address scalability issues is to leverage clustering algorithms to reduce communication and computation overheads \cite{8306880,dorri2017lsb}.
For example, Novo \cite{8306880} proposes a scalable blockchain solution for IoT systems. At the cost of additional communication delay, the proposed solution relies on a management hub that handles a group of IoT devices, thus reducing the number of interactions between things and the blockchain, effectively producing a scalable blockchain design.
A similar approach is instead proposed by Dorri \emph{et al.} \cite{dorri2017lsb}, where the authors design a scalable secure and privacy-preserving blockchain for IoT applications. 


  \begin{figure}[t]
\centering
    \centering
    \includegraphics[width=0.7\columnwidth]{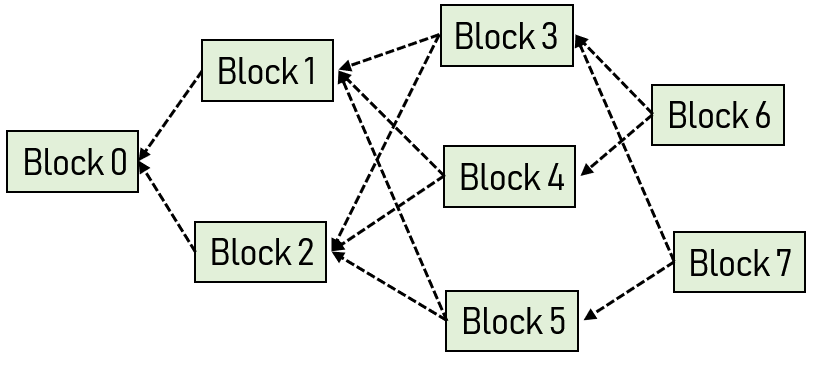}
    \caption{\label{fig:tangle} DAG blockchain (or Tangle).}
  \end{figure}
  \begin{figure}[t]
  \centering
    \includegraphics[width=0.7\columnwidth]{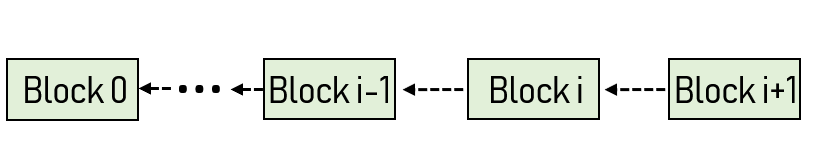}
    \caption{\label{fig:linear} Traditional linear blockchain}
\end{figure}  

Other approaches choose to revisit the structure of consensus mechanisms and the blockchain itself to provide ad-hoc solutions for the IoT. Glaring examples are the crypto-currencies IoT Chain (ITC) \cite{ITC} and IOTA \cite{popov2016tangle}. These currencies are built to provide lightweight blockchain technologies for the IoT. Specifically, together with other coins such as Byteball \cite{byteball}, ITC and IOTA aim at reshaping the linear structure of the blockchain to obtain a \textit{tangled} \cite{popov2016tangle} network represented by a direct acyclic graph (DAG).  The differences between traditional (linear) blockchain approaches and the DAG-based one are shown in Figs. \ref{fig:tangle} and \ref{fig:linear}.
In the DAG-based architecture (also called \textit{tangle}), blocks represents vertices of the DAG and edges are used to validate transactions. Specifically, to be included in the DAG, each new transaction $A$ must approve any two transactions $B$ and $C$ already included in the DAG. Approval of transactions is represented through directed edges going from a transaction to another. Accordingly, when $A$ is included in the DAG, it automatically generates two edges $A\rightarrow B$ and $A\rightarrow C$ that extend the DAG incrementally. The synergic usage of DAGs and blockchain technologies allows to dispose of the linear structure of traditional blockchains, facilitate transaction verification times and eliminate the need for mining as transactions are in charge of validating other transactions.
 
\subsection{IoT-tailored Security and Reliability}

Being able to remotely access one or more devices, together with the possibility to let them communicate and coordinate with each other autonomously is with no doubt a very useful and remarkable feature. However, this inevitably poses several concerns from the security and reliability viewpoints \cite{restuccia2018securing,lin2017survey}. 
The IoT is vulnerable to a wide variety of network attacks that undermine the confidentiality, integrity, authentication and availability. These aspects are fundamental requirements of any modern communication network and a variety of solutions for have been proposed in the literature \cite{restuccia2018securing,lin2017survey,alrawais2017fog,sadeghi2015security,granjal2015security}.
These surveys provide an exhaustive literature review of already existing solutions to design secure and reliable IoT systems. At the same, however, they show that many security solutions are not general enough and require ad-hoc solutions that involves new algorithms and software.

The blockchain already implements several mechanisms such as public/private encryption, hashing, consensus and fault-tolerance whose effectiveness in terms of security has been widely investigated and verified for many networking scenarios. For this reason, the blockchain has been identified as a pivotal technology to design secure and reliable IoT systems \cite{dorri2017blockchain,dorri2017lsb,banerjee2018blockchain,alphand2018iotchain}. 

\begin{itemize}
    \item \textit{Confidentiality:} data confidentiality is achieved when a given information (\textit{e.g.,} sensing data, transaction) can be accessed by intended devices only. In this context, public key encryption used to perform transactions in the blockchain can be seamlessly used to encrypt communications and data to be stored, thus effectively achieving confidentiality;
    \item \textit{Integrity:} to guarantee that data accessed and stored by IoT devices is reliable, integrity of contents must be ensured at all times. Again, the blockchain comes in help by providing useful mechanisms to guarantee integrity of data. Recall that the integrity of each block in the blockchain is verified by computing its hash value, and that the hash value of any block depends on the hash of previous blocks. Accordingly, not only hashing in the blockchain ensures integrity of a new block, but it also extends the integrity check to all previous blocks. As shown in \cite{dorri2017blockchain}, the same concept can be used in blockchain-based IoT networks to check the integrity of sensing data, transmitted data and transactions among devices and users;
    \item \textit{Authentication and Non-repudiation:} by leveraging on the already embedded public key encryption it is possible to implement signature-based security mechanisms, which are well-known to jointly provide authentication and non-repudiation \cite{zhou1999securing}. Recall that any node $B$ possessing the public key of a given device $A$ can i) decode messages encrypted by using $A$'s private key; and ii) encrypt messages with the public key of $A$. Since the private key of $A$ is known to $A$ only, public key encryption makes it possible to use private keys to generate an electronic signature of $A$. This signature is used to authenticate $A$ when it communicates with other nodes (each node can verify the signature by using $A$'s public key); and can be used for non-repudiation purposes to sign all the transactions included in the blockchain that involve $A$, thus effectively proving $A$'s activity on the blockchain.
\end{itemize}

\section{Conclusions}

In this paper, we have provided an overview of existing literature on the topic of blockchain for IoT, and presented a roadmap of research challenges that will need to be addressed to enable the usage of blockchain technologies in the IoT. First, we have briefly introduced the concept of blockchain in Section \ref{sec:blockchain}, followed by an overview of existing blockchain-based IoT applications in Section \ref{sec:iot_block_app}. Then, we have presented the major blockchain technologies for the IoT in Section  \ref{sec:techsforiot}. We have concluded the paper by discussing several research challenges in Section \ref{sec:res_challenges}. 

\footnotesize

\bibliographystyle{IEEEtran}

 \end{document}